# Stabilization of sawtooth instability by short gas pulse injection in ADITYA-U tokamak

Suman Dolui[1,2,*] Kaushlender Singh,[1,2] Bharat Hegde,[1,2] T. Macwan,[3] SK Injamul Hoque,[1,2] Umesh Nagora,[1,2] Jaya Kumar A.,[4] S. Purohit,[1] A. N. Adhiya,[1] K. A. Jadeja,[1] Harshita Raj,[1,2] Ankit Kumar,[1,2] Ashok K. Kumawat,[1,2] Suman Aich,[1,2] Rohit Kumar,[1] K. M. Patel,[1] P. Gautam,[1] Sharvil Patel,[1] N. Yadava,[5] N. Ramaiya,[1] M. K. Gupta,[1] S. K. Pathak,[1,2] M. B. Chowdhuri,[1] S. Sharma,[1,2] A. Kuley,[4] R. L. Tanna,[1] P. K. Chattopadhyay,[1,2] A. Sen,[1,2] Y. C. Saxena,[1,2] R. Pal,[6] and Joydeep Ghosh[1,2,†]

[1]*Institute for Plasma Research*, Gandhinagar 382428, India
[2]*Homi Bhabha National Institute*, Anushaktinagar, Mumbai 400094, India
[3]*Department of Physics and Astronomy, University of California Los Angeles, Los Angeles*, California 90095, USA
[4]*Department of Physics, Indian Institute of Science, Bangalore* 560012, India
[5]*Oak Ridge Associated Universities*, Oak Ridge, Tennessee 37831, USA
[6]*Saha Institute of Nuclear Physics*, Kolkata 700064, India

Experiments on ADITYA-U tokamak show a marked enhancement in the sawtooth period by application of short gas puffs of fuel that cause a modification of the radial density profile. A consequent suppression of the trapped electron modes then leads to an increase in the core electron temperature. This slows down the heat propagation following a sawtooth crash causing a delay in achieving the critical temperature gradient inside the $q = 1$ surface required for the next sawtooth crash to happen. The overall scenario has strong similarities with the behavior of sawtooth under electron cyclotron resonance heating (ECRH). Our findings suggest an alternate technique for sawtooth control that may be usefully employed in small- and/or medium-sized tokamaks that do not have an ECRH or any other auxiliary heating facility.

## I. INTRODUCTION

Sawtooth (swt) oscillation in a tokamak plasma is a quasiperiodic phenomenon where electron temperature gradually rises in the core region followed by a sudden crash via internal disruption [1,2]. The phenomenon is detrimental to heating and particle confinement in the core of the plasma. The swt crash also leads to generation of instabilities affecting the steady-state scenario [3,4] of fusion-grade plasmas. Therefore, the disruption needs to be controlled; however, its complete suppression is not desirable, as it can lead to impurity peaking, which causes substantial degradation of confinement [5]. A controlled stabilization is thus ideally desirable to proceed toward the fusion reactor goal. The swt crash is attributed to magnetic relaxation by Kadomtsev's well-known resistive reconnection model [6], which considers excitation and nonlinear growth of an $m/n = 1/1$ internal kink or tearing mode near the $q = 1$ surface ($m$ and $n$ are poloidal and toroidal mode numbers, respectively, and $q$, the safety factor, is a measure of the helicity of the magnetic field) to be the cause behind the internal disruption. The model predicts crash times that agree well with several tokamaks, but predicts much larger crash time than that observed (∼100 μS) in large tokamaks. Furthermore, the mode oscillations are not always observed before the crash. So, understanding swt relaxation remains an open question and deserves further studies to control it effectively. Several models have been proposed to explore possible mechanisms for the stabilization: some are based on controlling the fast particles generated during heating mechanisms to lower the pressure gradient, while others depend on decreasing the magnetic shear near the $q = 1$ region [7]. They have conjectured that the growth rate of the $m/n = 1/1$ mode can be influenced by the presence of energetic ions [8,9], by intrinsic plasma rotation [10], or by shear in rotation [9,11]. Dynamic modification of plasma parameters near $q = 1$ surface seems to be crucial to prevent the crash. Experimentally, localized heating by lower-hybrid current drive, ion-cyclotron resonance heating, electron cyclotron resonance (ECR) heating, and neutral beam heating [8,12] have demonstrated substantial control of the swt destabilization. Recent experiment in DIII-D tokamak [13] using ECR heating suggests that the temperature turbulence inside the inversion radius (a radius beyond which the swt character reverses) is involved in triggering the swt crash. In this paper, we present employment of a successful alternate scheme that experimentally demonstrates delaying the swt crash substantially in ADITYA-U tokamak, by controlling the plasma temperature profile inside the $q = 1$ surface. The scheme involves injection of short puffs of fuel gas (containing ∼$10^{17}$–$10^{18}$ molecules/m$^3$) in the plasma edge. The injection results in a cold pulse propagation [14–16], which modifies the radial profile of plasma density in the mid-radius region and subsequently modifies the electron temperature profile in the core region (inside $q = 1$ surface). The observed swt stabilization results closely resemble those obtained through localized ECR heating near the $q = 1$ surface in DIII-D tokamak [17]. So, the present mechanism provides a relatively simpler technique for studying swt phenomena systematically in tokamaks.

*Contact author: suman.dolui@ipr.res.in
†Contact author: jghosh@ipr.res.in

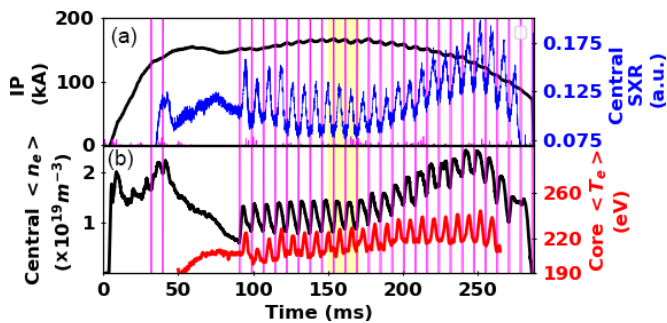

FIG. 1. Temporal evolution of plasma parameters (Shot No. of 33689): (a) $I_P$ (black), SXR emission (blue) and (b) central chord–integrated density $n_e$ (black), core $T_e$ (red).

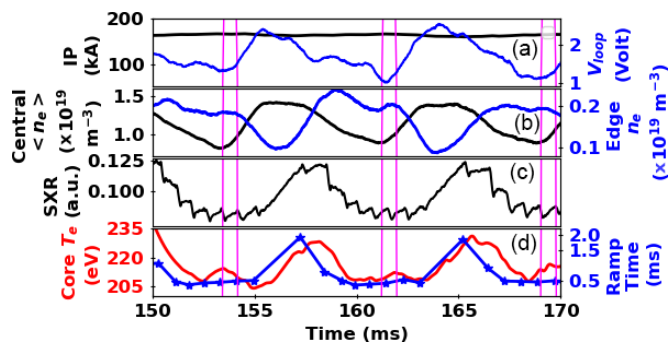

FIG. 2. Temporal evolution of (a) $I_P$ (black), $V_{loop}$ (blue); (b) $\langle n_e \rangle$ (black), edge $n_e$ (blue); (c) central chord SXR intensity (black); and (d) core $T_e$ (red solid line), ramp time of sawtooth cycle (blue stars) (Shot No. of 33689).

## II. EXPERIMENT

The present experiment is conducted in the Ohmically heated hydrogen discharges of the ADITYA-U tokamak [18] with toroidal magnetic field $B_\Phi = 1.2–1.4$ T and plasma current $I_P = 150$-$180$ kA. The plasma major radius is $R = 0.75$ m and a toroidal-belt limiter defines the minor radius $a$ to be 0.25 m. The central chord–averaged electron density ($\bar{n}_e$) and electron temperature ($T_e$) are $1$–$4 \times 10^{19}$ m$^{-3}$ and 200–350 eV, respectively. The radial density profile is measured using a seven-channel (3 heterodyne +4 homodyne) microwave interferometer [19], whereas the core electron temperature is obtained by soft-x-ray (SXR) emission measurements using a foil-ratio method [20]. An SXR tomography camera, consisting of 16-channel AXUV (Absolute extreme ultra-violet) photodiodes, scans the radius from $r = -10$ to 10 cm [21]. The edge density and temperature are measured by Langmuir probes [22]. The locations of various diagnostics and gas-puff injection point are described in Ref. [14]. Figure 1 shows a typical discharge evolution of the parameters $I_P$, loop-voltage $V_{loop}$, central chord–averaged SXR emission, and density $\bar{n}_e$ with the application of periodic hydrogen/deuterium gas puffs. The amount of gas injection is controlled by varying the voltage pulse width to the piezo-valve; a pulse width of $\sim$1 ms injects $\sim$10$^{17}$ molecules [14]. The swt activities are observed in various parameters including the SXR emission from the central chord [Fig. 1(a)].

## III. RESULTS

### A. Sawtooth period enhancement

The swt cycle shows a quiescent ramp phase ($\sim$600 μS), followed by a crash ($\sim$50−100 μS). The peak corresponds to $\sim 10\%$ rise in core plasma temperature $T_{e0}$. The swt inversion radius is $r_{inv} \sim 4$−$6$ cm. Taking $T_{e0} \sim 300$ eV and core density $n_{e0} \sim 2 \times 10^{19}$ m$^{-3}$ in these discharges, the observed swt crash time is close to the characteristic Sweet-Parker timescale $\sqrt{\tau_R \tau_A} \sim 85$ μS, where $\tau_A = r_{inv}\sqrt{4\pi\rho_m}B_\varphi^{-1}$, $\tau_R = 4\pi r_{inv}^2/\eta c^2$, $\rho_m$ is the mass density, and $\eta$ is the plasma resistivity.

To examine the swt activity closely, the time span 150–170 ms during the plasma current flat-top is expanded in Fig. 2. Impact of the gas pulse injection (done repeatedly at 8 ms intervals) is clearly visible on the swt oscillations in SXR intensity [Fig. 2(c)], as well as on $I_P$, $V_{loop}$, $\bar{n}_e$, $n_e^{edge}$ (at $r = 24$ cm), and $T_{e0}$. The most interesting point to note here is that the ramp phase is extended substantially after each gas puff and the crash is delayed. The ramp time of each cycle is plotted in Fig. 2(d). After each gas puff, the ramp time is increased to nearly 1–1.5 ms, which is about double the time compared to that seen without gas puff. The increment in the ramp time gradually decreases in subsequent swt pulses and attains the pre-gas-puff value after three to four cycles depending on the amount of gas injected. The results are confirmed by repeated observations over several tens of discharges covering hundreds of swt cycles. The increment in swt ramp time is seen to rise proportional to the amount of gas injected, until it affects the plasma adversely (Fig. 3) [23].

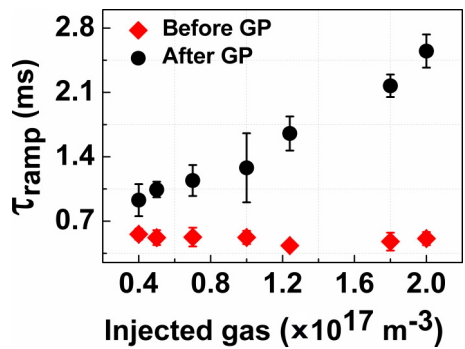

FIG. 3. Ramp time of swt cycle (black circle) with amount of injected gas. Corresponding ramp time in the absence of gas puff (red triangle).

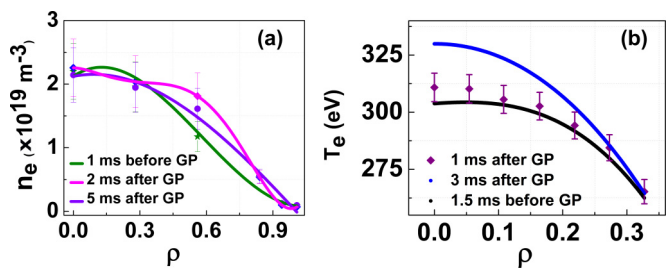

FIG. 4. Radial profiles of (a) density and (b) core $T_e$ before and after gas puff. The curves are the spline fits to the experimental data (symbol).

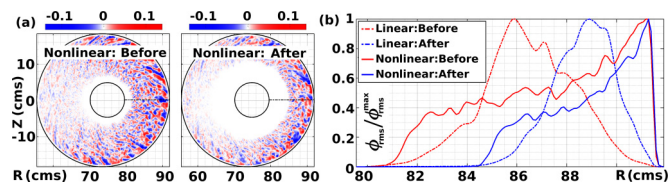

FIG. 5. (a) Electrostatic perturbed potential on the poloidal plane during the nonlinear phase and (b) the corresponding radial variation of the rms electrostatic potential normalized with the maximum values during the linear and nonlinear phases.

### B. Modification of radial profiles of density and core temperature

As shown in Fig. 2(b), following the gas-pulse injection, $\bar{n}_e$ increases sharply and reaches a peak in about 1 ms after the injected pulse, whereas the edge plasma density (at $\rho = r/a \sim 0.8 - 1$) decreases. The temperature of the core plasma also increases [Fig. 2(d)]; however, it starts rising later, near about the time the density enhancement reaches its peak. The temperature rise time to the peak is 2–3 ms, which is smaller than the energy confinement time ($\sim 5$ ms). The above observations clearly suggest a cold pulse propagation phenomenon by the gas puff. The SXR emission remains confined within the core region of the plasma ($\rho < 0.3$ in this experiment), where the electron temperature is more than 100–200 eV. The SXR emission intensity is known to be proportional to $n_e^2$, but more sensitive to $T_e$. In the experiment, the temporal evolution of the gas puff–induced SXR emission [Fig. 2(c)] closely follows the core temperature variation [Fig. 2(d)], rather than the density variation [Fig. 2(b)]. This clearly indicates that the gas puff affects the temperature in the core region, where the density does not change much. Experimental results shown in Fig. 4 are consistent with this proposition. In Fig. 4(a), we compare the radial profiles of density with and without gas puff. After gas injection, the density does not change much inside, but increases considerably outside the core ($\rho > 0.3$) of the plasma. In other words, the density profile is flattened by the gas puff. The flattened density profile reduces the density fluctuation and the plasma energy also increases [14]. The maximum density increment is observed at about 1 ms after the gas injection between $\rho = 0.4$ and 0.6, indicating the density to change mainly outside the SXR emission zone.

In contrast, radial temperature profiles [Fig. 4(b)] show the temperature to rise in the core region. The sign inversion of temperature increment occurs in the vicinity of the $q = 1$ surface, pointing toward the profile to become more peaked after the gas puff. The above observations resemble those observed in the cold pulse propagation phenomenon in DIII-D and ASDEX Upgrade tokamaks caused by impurity injection. They have been successfully explained by a model [15], which predicts that flattening of the electron density profile stabilizes the Trapped Electron Mode (TEM) instability, thereby leading to turbulence reduction [24,15] and reduced transport, if the heat transport is dominated by electron heat flux.

### C. TEM suppression

To identify whether electron- or ion-scale turbulences are involved here, self-consistent gyrokinetic simulations [25] are conducted for the present ADITYA-U discharges using the GTC code for cases before and after gas pulse injection, incorporating the equilibrium profiles obtained through IPR-EQ code [26]. The primary feature of the linear eigenmode structure, which propagates in the electron diamagnetic direction with $k_\perp \rho_i \sim 0.7$, is identified as a trapped electron-driven instability (TEM) [23,24,27], which plays a significant role in generating the anomalous turbulent transport. In linear regime, the TEM turbulence structure shows a relatively larger radial extent on the low-field side before gas injection [23], whereas with gas puff it contracts with the turbulence peak moving outward [shown in Fig. 5(b)]. In nonlinear regime, however, the turbulence encompasses almost the entire plasma ($\rho = 0.1-1$) with no gas injection, whereas with gas injection it is suppressed in a broad region (up to $\rho \sim 0.25$, i.e., near the $q = 1$ surface) of the core plasma [Fig. 5(b)] and it is clearly depicted in the poloidal plane plot of Fig. 5(a). It is plausible that such fluctuation suppression with gas puff causes reduction of heat transport from a broader region, and is likely to produce gradual rise of core temperature for a longer time to a significant level. This is manifested also in the observed temperature profile of Fig. 4(b). The observations are very much like those obtained with temperature modification experiments near the $q = 1$ surface by ECR heating [12]. The swt cycle before gas puff shows a gradual increase in the SXR emission until a growing precursor oscillation (oscillation period $\sim 50~\mu S$) develops on it followed by a crash. After gas puff, similar precursor oscillation develops, but not before a substantial delay.

### D. Identification of sawtooth precursor mode

The SXR emissions along two chords on either side of the center are plotted in Fig. 6. In both the cases, the precursor oscillation grows in amplitude until the crash; however, the oscillation survives the crash and stays at a steady level [Fig. 6(d)]. Detailed analysis of the oscillation [Fig. 6(c)] distinctly shows $180°$ phase difference between them indicating an odd $m$ mode (most likely $m = 1$). In our experiments, taking the plasma parameters of the core, the ideal $m = 1$ kink mode is considered stable [Lundsquist number $(S) \sim 10^5$] [28]. Hence, the observed $m = 1$ oscillation may be due to excitation of a resistive mode producing a slowly growing island. The survival of the oscillations supports partial reconnection during the swt phase. Incomplete reconnection has been described by several models [29]. These include a flattened $q$ profile, shear flow, stochastic magnetic fields [30], diamagnetic suppression of the nonlinear internal kink mode [7], pressure effects at the magnetic island boundary, etc.

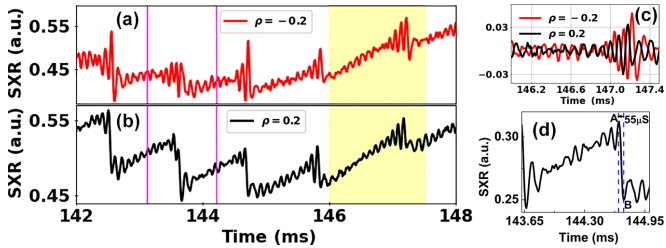

FIG. 6. SXR intensity at (a) $\rho = 0.20$ and (b) $\rho = -0.20$. (c) SXR fluctuations at $\rho = 0.20$ (black) and $\rho = -0.20$ (red). (d) A typical swt crash phase with presence of pre- and postcursor oscillations.

The Kadomtsev model can provide a possible explanation of the swt-period enhancement with gas puff as follows: The observed increase of the core temperature results in a delay in current penetration for the swt cycle initiated after the crash of the pre-gas-puff cycle, and it delays the onset of the $m = 1$ mode [31]. Moreover, the growth rate of the $m = 1$ mode decreases as the core temperature increases. A combination of these two effects may result in the swt-period extension. However, the amplitude of the $m = 1$ oscillations at the crash time is observed to be random irrespective of whether the crash occurs with or without gas pulse. The persistence of these oscillations observed in Fig. 6(d) at the onset as well as after the crash indicates that complete resistive reconnection may not be related to the internal disruption in our scenario [32]. Moreover, with gas puff–induced $\Delta T_e = 30$-50 eV, the decrease in mode growth rate would lead to a delay of only $\sim 1\,\mu S$ in critical island formation. Also, the change in core $q$ value ($\sim 0.007$-$0.01$) is negligible [32] within the period of elevated temperature after each gas puff. In the present experiment, the swt inversion takes place around the radius $\rho \sim 0.2$ (the $q = 1$ surface), regardless of the presence or absence of the gas puff (see Fig. 8). As the density remains flat inside this radius, the temperature profile determines the pressure profile there.

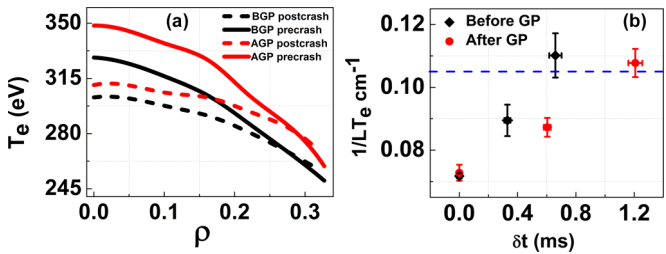

FIG. 7. (a) $T_e$ profile at the core at pre- and post-swt crash time with and without gas pulse injection. (b) The threshold of $1/L_{T_e}^{\text{core}}$ value at $\rho \sim 0.2$ to occur an swt crash.

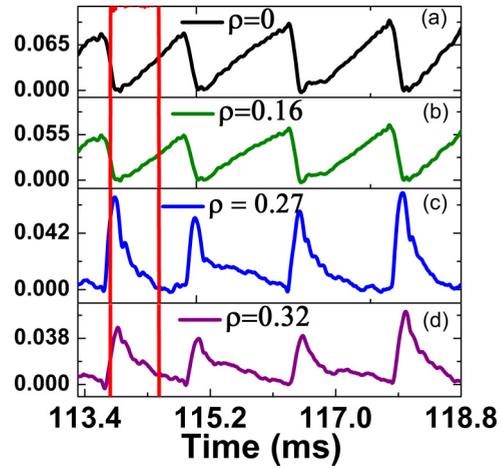

FIG. 8. SXR intensity time traces from different chords, $\rho = 0$, 0.16, 0.27, and 0.32, showing swt and inverted swt (vertical lines indicate time of gas-puff pulse).

### E. Role of critical core temperature gradient in sawtooth crash

The $T_e$ profiles just before and after the crash are shown in Fig. 7(a). The profile flattened after the crash in both cases. To investigate the change in inverse $T_e$ scale length, $1/L_{T_e}^{\text{core}} = -(1/T_e)(dT_e/dr)$ with time at $\rho \sim 0.2$, $1/L_{T_e}^{\text{core}}$ is plotted in Fig. 7(b) at three different time intervals from the beginning to the end of the swt period, for both with and without gas-puff cases. Each data point in the figure represents an average of tens of swt cycles, with error bars representing the scatter in the measurements.

Interestingly, swt crashes take place around $1/L_{T_e}^{\text{core}} \sim 0.11$ for all cycles, regardless of gas puff being present or not. The observation strongly points toward a threshold for the temperature gradient at the inversion radius that relates to the crash. The Ohmic heating and heat loss rates determine the evolution of $T_e$ profile near the inversion radius during the ramp phase of a new swt cycle. The current diffusion time in the core area is substantially long ($\sim 30$ ms). So, the Ohmic power remains almost unchanged over swt cycles before and after gas puff. Then, it is the heat transport from the core region that determines the temperature gradient.

The reduced heat diffusivity with the gas puff is experimentally demonstrated in Fig. 8, which displays the time traces of SXR intensity for chords $\rho = 0$, 0.16, 0.27, and 0.32. The observed SXR emission inverses its character outside $\rho \sim 0.2$ (the inversion radius); i.e., after an swt crash inside, it increases sharply outside. This indicates sudden heat pulse propagation outward. The SXR intensity returns to the precrash value within 500 $\mu S$ for the case of no gas injection, whereas it takes longer time $\sim 1000\,\mu S$ with gas puff [Figs. 8(c) and 8(d)], suggesting reduced heat diffusivity. A plausible explanation is that the reduction in TEM-induced heat diffusivity in the core region by gas puff delays the attainment of the critical $T_e$ gradient, thereby prolonging the swt ramp time.

From Fig. 7(b), it is evident that the $1/L_{T_e}^{\text{core}}$ increases gradually during the rising phase of an swt cycle and attains its maximum value just before the crash. In Fig. 9(a), the amplitude of the SXR intensity fluctuation is plotted against $1/L_{T_e}^{\text{core}}$ at $\rho \sim 0.2$. As the $T_e$ profile steepens, i.e., as $1/L_{T_e}^{\text{core}}$ increases, the fluctuation amplitude is seen to increase slowly. However, at a critical value of $1/L_{T_e}^{\text{core}}$, the amplitude increases sharply and the crash follows. Interestingly, continuous morlet2 wavelet transform analysis [33] of the central chord SXR intensity [Fig. 9(b)] indicates development of a broadband turbulence in the core temperature before the crash, which reduces considerably after the crash (vertical magenta lines). It is to be noted here that, as the spectra might be influenced by the swt nature itself, Fig. 9(b) is obtained by removing the contribution of the spectra of a fluctuation-free swt cycle of similar amplitude and timescale generated synthetically. So, the data strongly indicate that the swt crash may be triggered by development of a broadband turbulence in core $T_e$ when a critical value of $1/L_{T_e}^{\text{core}}$ is reached [13,24] at the $q = 1$ surface.

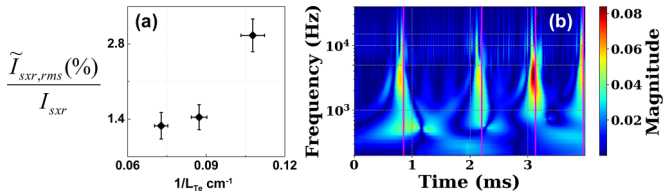

FIG. 9. (a) RMS value of SXR intensity fluctuation as a function of inverse $T_e$ scale length at $\rho = 0.2$. (b) Continuous morlet2 wavelet transform analysis of the central chord SXR intensity.

## IV. CONCLUSION

In conclusion, the present experiments in ADITYA-U tokamak demonstrated that edge-injected short gas-puff pulse could be a novel way of controlling swt relaxation. Without gas puff, the swt ramp crashes shortly following the appearance of the $m = 1$ mode, though the oscillations last even afterward. After a gas puff, outside of the core region undergoes an electron density rise within 1 ms, flattening the density profile, which suppresses the TEM modes and reduces heat transport. In contrast, it raises the core temperature later within the $q = 1$ surface ($\rho \sim 0.20$) making the profile steeper. The swt crash is seen to occur when a critical temperature gradient ($1/L_{T_e}^{\text{core}} \sim 0.11$) is reached near the inversion radius ($\rho \sim 0.2$) irrespective of gas puff [25]. A sharp rise of a broadband turbulence at this gradient seems relevant to the swt crash. Reduced heat transport with gas injection delays attaining this critical gradient. The observed gas puff–induced variation of temperature profile influences the swt cycle resembling those observed due to localized ECR heating experiments near $q = 1$ surface. This new technique, therefore, provides an alternative and relatively simpler method for in-depth study of swt phenomena in tokamaks without facing the complexities of external heating.

## DATA AVAILABILITY

The data that support the findings of this article are not publicly available. The data are available from the authors upon reasonable request.